 \documentclass[12pt,preprint]{aastex}

\usepackage {epsf}
\slugcomment{draft}

\begin{document}

\title{The Non--Monotonic Dependence of Supernova and Remnant Formation on 
Progenitor Rotation}

\author{Shizuka Akiyama, J. Craig Wheeler}
\affil{Astronomy Department, University of Texas, Austin, TX 78712}
\email{shizuka@astro.as.utexas.edu,wheel@astro.as.utexas.edu}

\begin{abstract}

Traditional models of core collapse suggest the issue of successful versus
failed supernova explosions and neutron star versus black hole formation
depends monotonically on the mass (and metallicity) of the progenitor star
with mass above some cutoff leading to black holes with or without
attendant supernova explosions. Here we argue that the issue of success or
failure of the explosion or other possible outcomes may depend
non--monotonically on the rotation of the progenitor star even at fixed
progenitor mass and composition.  We have computed ``shellular" models of
core collapse for a star of 15 M$_{\odot}$ with initial central angular
velocity, $\Omega_0$, in the range 0.1 -- 8 rad s$^{-1}$ until a few
hundred ms after bounce to explore qualitative trends.  The model with
$\Omega_0$ = 4 rad s$^{-1}$ gives the maximum post--bounce rotation with
rotation comparable to that necessary for secular or dynamical instability
to occur. Models with $\Omega_0$ $>$ 4 rad s$^{-1}$ bounce at sub--nuclear
density with $\gamma \sim 4/3$ and subsequently slowly contract.  The
non--monotonic behavior will be manifested in the rotation of the
proto--neutron star and hence in the strength of the associated magnetic
field that will be generated by shear in that rotating environment.  We
estimate that our maximally rotating and shearing models generate toroidal
fields approaching or exceeding $10^{17}$ G, strengths nearing dynamical
significance.  One implication of this nonmonotonicity is that the process
of rotating, magnetic core collapse may itself provide a filter to select
specific outcomes from the distribution of initial rotation states. A
possible outcome is black hole -- neutron star -- black hole with
increasing initial angular momentum of the progenitor. Within the regime
of successful explosions leaving neutron stars behind, a subset may spin
rapidly enough, either initially or during the subsequent
de--leptonization contraction phase, to drive an $\alpha - \Omega$ dynamo
and hence produce the large dipole field associated with magnetars.

\end{abstract}


\section{Introduction}

The collapse of the core of a massive star can lead to a number of
possible outcomes: an explosion leaving behind a neutron star, frequently
observationally manifested as a pulsar; an explosion leaving behind an
especially strongly magnetized neutron star, a magnetar
\citep{dun92,kou99}; a black hole associated with an otherwise successful
explosion; or complete collapse to a black hole with little or no mass
ejection or light output.  Classical Type II supernovae may be one extreme
of this distribution of properties, gamma--ray bursts another.

Traditional, spherical or nearly spherically--symmetric models of core
collapse have led to the widespread assumption that the outcome of core
collapse depends monotonically on the mass of the progenitor star.  The
assumption, implicit or explicit, is that with increasing binding energy
and overburden in more massive stars, the explosion becomes more difficult
and hence black hole formation though fall--back or complete collapse
becomes more likely \citep{bur88,fry99,fry01}.

The physics that dictates the reversal of core collapse into an explosion
is not yet fully understood.  Neutrino processes undoubtedly play a major
role \citep{ram00, ram02, lie01a, lie01b, mez01, tho05}, but the discovery
of ubiquitous asymmetries in core collapse supernovae \citep{wan96, wan01,
wan02, wan03, leo00, leo01, leo02} and the success of jet--induced models
to form such asymmetries \citep{kho99, kho01, hof01} and effects on the
nucleosynthesis \citep{maeda03} have brought new focus to the question of
whether rotation \citep{FH00, mul04, fry04, ott04} and magnetic fields
\citep{whe00, whe02, kot04, yam04} play an important part in the explosion
process.  In particular, \citet{aki03} showed that the magnetorotational
instability \citep[MRI;][]{BH98} will be a ubiquitous effect in the
differentially--rotating structure associated with a newly formed neutron
star where predominantly toroidal fields of $\sim 10^{15} - 10^{16}$ G are
expected. \citet{tho05} explored the possibility that heating from
magnetic dissipation may boost that from neutrino deposition.

In general, more rapid rotation of the new--born neutron star will lead to
greater shear and to production of larger magnetic fields, whatever their
influence on the outcome of the explosion.  On the other hand, given the
plethora of single and binary star evolutionary paths that can influence
the progenitor rotational evolution, it is difficult to believe that all
stars evolve to just the right rapidly rotating pre--collapse state
\citep{lan04}.  There must be a distribution of initial angular momenta of
the progenitor stars, of the pre--collapse cores, and of the neutron
stars.  Here we argue that the processes of collapse, rotation, shear, and
magnetic field generation may provide the filter to select specific
outcomes from this distribution of initial rotation states. In particular,
we argue that the final angular velocity that determines the shear and
hence the strength of the magnetic field generated by the MRI will not
vary monotonically with the initial rotation speed of the pre--collapse
core.  This, in turn, suggests that the outcome of core collapse at a
single mass may depend in a non--monotonic manner on the initial angular
momentum of the pre--collapse core.

In \S 2 we outline the basic nature of the argument and present a simple
analytic model to show the general trend. In \S 3 we present numerical
results from a ``shellular" collapse with a centrifugal potential to
extend the argument.  Section 4 gives a summary and conclusions.

\section{Qualitative Non--monotonic Behavior and a Simple Model}

The basic thrust of our argument is that the final outcome of core
collapse will naturally lead to a non--monotonic behavior of the neutron
star rotational velocity, the shear, and the magnetic field with the
initial angular momentum of the core. This is because for low initial
angular momentum the outcome will be very similar to that for
non--rotating models, whatever that is. As the initial angular momentum
increases, so will the resulting rotation rate of the neutron star.  If
the initial angular momentum increases sufficiently, however, centrifugal
effects will prevent the neutron star from becoming as compact (at least
along the equator) and hence the angular velocity of the neutron star will
be less. The result is that the angular velocity and hence the shear and
associated magnetic field may have a distinct maximum as the initial
angular momentum of the pre--collapse core is increased.  This could
single out a rather narrow range of initial angular momenta at which the
angular velocity, shear, and resulting magnetic field will be a maximum.

The basic behavior can be illustrated with a simple model. If we treat the
effects of rotation as an effective centrifugal potential, then pressure
gradients in the post--bounce quasi--static proto--neutron star core can
be described by:
\begin{equation}
\label{grad-P}
\nabla P = - \rho \nabla \Phi_G (1 - \epsilon),
\end{equation}
where 
\begin{equation}
\epsilon = \left|\frac{\Phi_c}{\Phi_G}\right| \sim \frac{L^2}{GM^3R}
\end{equation}
is the absolute value of the ratio of the centrifugal potential, $\Phi_c$,
to the gravitational potential and $L$ is the initial angular momentum of
the pre--collapse core which is assumed to be conserved, so that $L \sim
MR^2\Omega$ where $M$ is the mass of the neutron star, $R$ its radius,
and $\Omega$ its angular velocity.

To simplify the argument, we consider eq. (\ref{grad-P}) in terms of
characteristic quantities and write,
\begin{equation}
\label{P-sim}
P \sim \rho \Phi_G (1 - \epsilon).
\end{equation} 

If we consider a simple polytropic equation of state, $P = K \rho^{\gamma}$,
then with $\rho \sim M/R^3$ in eq. (\ref{P-sim}) we find
\begin{equation}
\label{R1}
\left(\frac{R}{R_0}\right)^{3\gamma - 4} \sim (1 - \epsilon)^{-1},
\end{equation}
where $R_0$ is the radius of the neutron star in the absence of
rotation. Since $\Omega \sim L/MR(L)^2$, eq. (\ref{R1}) is
an implicit equation for $\Omega$ as a function of $L$.

To see the form of $\Omega(L)$, we assume that $\epsilon$ is
a small quantity. With the appropriate Taylor expansion we find
\begin{equation}
R^{3\gamma -3} - R_0^{3\gamma -4}R \sim \frac{R_0^{3\gamma -4}L^2}{GM^3}.
\end{equation}
If we further assume that $R = R_0 (1 + \eta)$ where $\eta$ is
another small quantity we find
\begin{equation}
\eta =  \frac{L^2}{3(\gamma - \frac{4}{3})GM^3R_0}.
\end{equation}  
Note that this factor by which the radius is amplified by
rotation can be exaggerated compared to the amplitude of the parameter
$\epsilon$ because of the factor of $\gamma - \frac{4}{3}$ in the
numerator. We will explore a related effect in the numerical
models presented below.
   
We then define a characteristic angular momentum, $L_c$, as
\begin{equation}
L_c = \left[3(\gamma - \frac{4}{3})GM^3R_0\right]^{1/2},
\end{equation}
where $L_c$ is related to, but not identical with, the Keplerian
angular momentum. With the definition of eq. 7, we can write
\begin{equation}
\label{rot}
\Omega \sim \frac{L}{MR_0^2\left[1 + 
\left(\frac{L}{L_c}\right)^2\right]^2}.
\end{equation}

We thus see that for $L \ll L_c$, $\Omega$ increases with L, but that as
$L$ becomes comparable to and then exceeds $L_c$, $\Omega$ will tend to
decrease with further increase in $L$.  Eq. \ref{rot} is not applicable in
the regime $L \gtrsim L_c$ because of the small parameter approximations
made, but the general trend is clear. To avoid these small parameter
approximations a numerical model is needed, to which we now turn.

\section{Time--Dependent Behavior with Varying Initial Angular 
Momentum} 

In order to demonstrate the simple model described above, we have
simulated the collapse of an iron core of a 15 M$_{\odot}$ progenitor
model with a one--dimensional flux--limited diffusion code \citep{ita87}.  
We assign an initial rotational velocity profile to the one--dimensional
initial iron core, and calculate the subsequent angular velocity by
assuming conservation of angular momentum in each mass shell.  The angular
momentum in the collapsing core can safely be assumed to be conserved
until core bounce according to the two--dimensional simulations of
\citet{yam04} (see their Fig. 11).  We have modified the original code to
include the centrifugal potential in accord with eqs. 1 and 2 in order to
assess rotational feedback on the dynamics.  Specifically, we have added a
term $+{\rm r}\Omega^2$ to the gravitational acceleration, $-GM_{\rm
r}/{\rm r}^{2}$, in the momentum equation. We keep in mind that our
calculations presented here after bounce and with significant rotational 
energy are not realistic because of our assumptions of spherical symmetry
and conservation of angular momentum.  Nevertheless we have identified
some interesting trends that can be explored with multidimensional
simulations.

\begin{figure}[htp]
\centering
\includegraphics[totalheight=0.4\textheight]{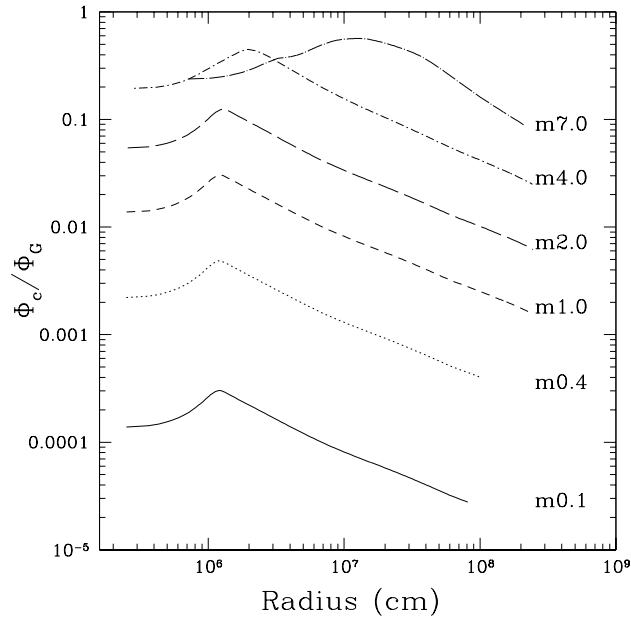}
\figcaption[centrifugal.eps]{The ratio of the centrifugal potential,
$\Phi_{\rm c}$, to the gravitational potential, $\Phi_{\rm G}$,
as a function of radius for selected models at bounce. 
\label{cent}}
\end{figure}

The rotational profile employed here is of the form \citep{MM89}
\begin{equation} 
\Omega(r) = \Omega_{0} \frac{R^{2}}{r^{2} + R^{2}},
\end{equation}
where $R = 10^{8}$ cm, and $\Omega_{0}$ is the initial central value of
the rotational velocity profile.  We study the non--monotonic behavior of
post--collapse rotation by varying the value of $\Omega_{0}$.  A total of
14 models have been computed with values of $\Omega_{0}$ varying from 0.1
to 8.0 s$^{-1}$ (see Table 1).  The models are indicated by the value of
$\Omega_{0}$, i.e. m0.1 for the model with $\Omega_{0} = 0.1$ rad
s$^{-1}$.  The corresponding values of the ratio of rotational energy to
absolute value of the gravitational energy $T/|W|$ are given in Table 1
and range from 8.58 $\times 10^{-4}\%$ to 5.49$\%$.  Maclaurin spheroids
are unstable to secular instability when $T/|W| > 14\%$ and to dynamical
instability when $T/|W| > 27\%$ \citep{tas78}.  The value of $T/|W|$ for
the onset of secular instability in more realistic situations is estimated
to be comparable to the value for Maclaurin spheroids (Shapiro \&
Teukolsky 1983; however see \S 4 for a discussion of the bar mode
instability).  Therefore our initial rotational profiles are stable to the
secular instability even for our most rapidly rotating model.  The
amplitude of the centrifugal potential is small for the less rapidly
rotating models in relative terms compared to the gravitational potential
and the ambient pressure gradients.  As shown in Fig. \ref{cent},
characteristic values of the centrifugal potential for models with
$\Omega_0 \sim 1$ s$^{-1}$ at bounce are $\sim 10^{-3} - 10^{-2}$ of the
gravitational potential.  Our most rapidly rotating models approach or
exceed the conditions for instability after collapse, as will be
illustrated below.

\begin{deluxetable}{ccrrrrrrrrcrl}
\tabletypesize{\scriptsize}
\tablecaption{Parameters of the models calculated.}
\tablewidth{0pt}
\tablehead{
\colhead{Model} & \colhead{$\Omega_{0}$ (s$^{-1}$)} & 
\colhead{$T/|W| (\%)$}
}
\startdata
m0.1 & 0.1 & 8.58 $\times$ 10$^{-4}$\\
m0.2 & 0.2 & 3.43 $\times$ 10$^{-3}$\\
m0.4 & 0.4 & 1.37 $\times$ 10$^{-2}$\\
m0.5 & 0.5 & 2.15 $\times$ 10$^{-2}$\\
m0.6 & 0.6 & 3.09 $\times$ 10$^{-2}$\\
m0.8 & 0.8 & 5.49 $\times$ 10$^{-2}$\\
m1.0 & 1.0 & 8.58 $\times$ 10$^{-2}$\\
m2.0 & 2.0 & 0.343\\
m3.0 & 3.0 & 0.773\\
m4.0 & 4.0 & 1.37\\
m5.0 & 5.0 & 2.15\\
m6.0 & 6.0 & 3.09\\
m7.0 & 7.0 & 4.21\\
m8.0 & 8.0 & 5.49\\
\enddata
\end{deluxetable}

\begin{figure}[htp]
\centering
\includegraphics[totalheight=0.4\textheight]{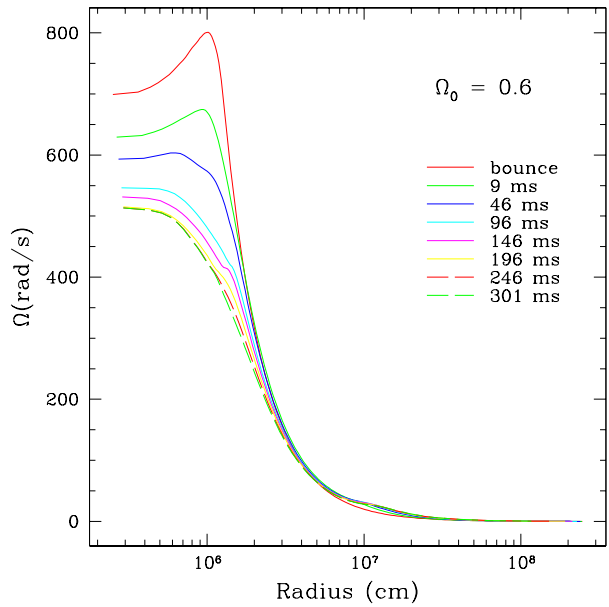}
\figcaption[omega0.6.eps]{Rotational profiles of the model with
$\Omega_{0} = 0.6 s^{-1}$ as a function of radius at elected times after
bounce.
\label{ome0.6}}
\end{figure}

\begin{figure}[htp]
\centering
\includegraphics[totalheight=0.4\textheight]{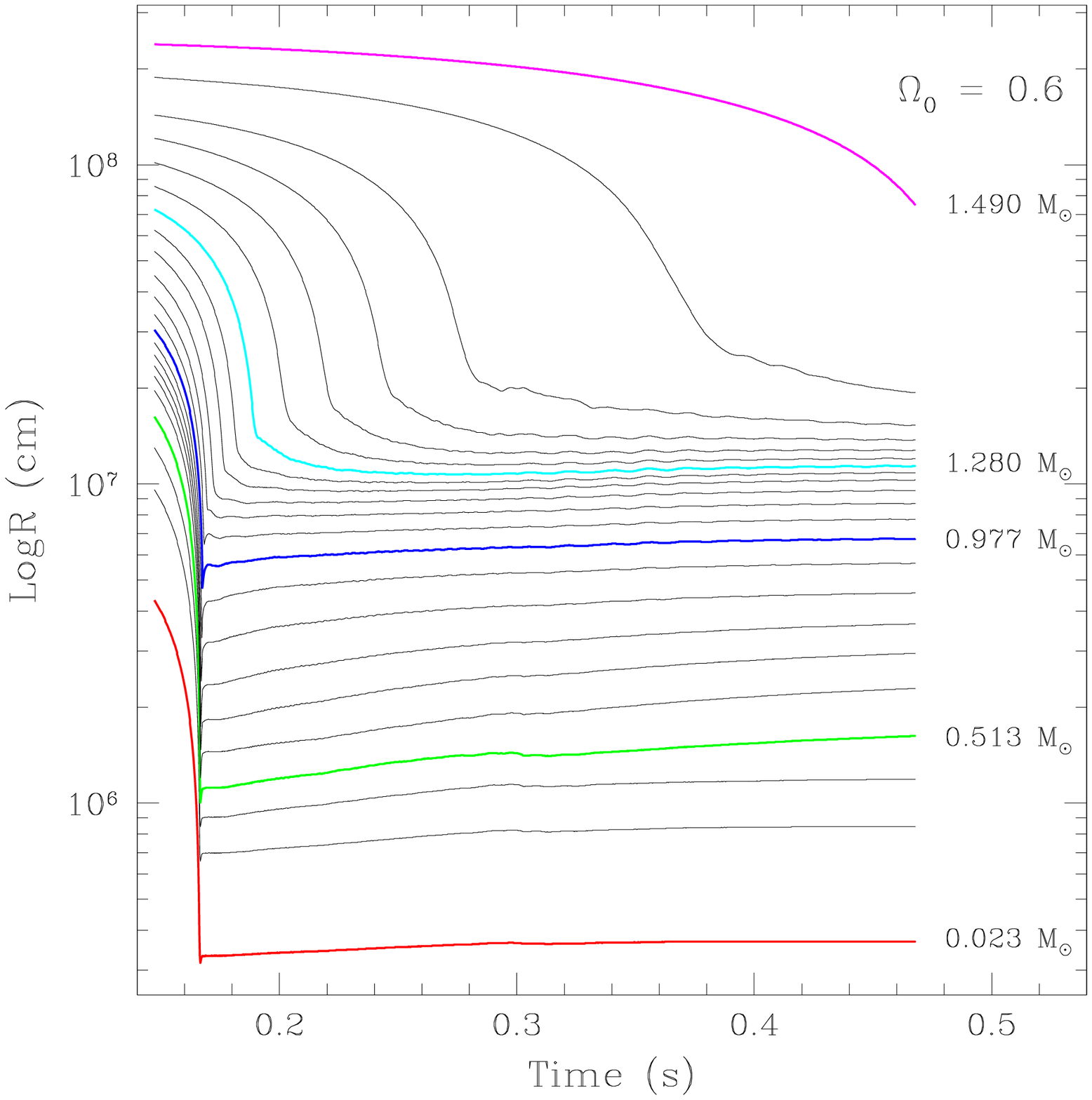}
\figcaption[stream0.6.eps]{The radius versus time of selected 
mass zones for model m0.6.  \label{st0.6}}
\end{figure}

For a model with modest initial rotation, the rotational velocity is most
rapid at the bounce, as the core is most compact then. Fig.  \ref{ome0.6}
shows the evolution of the angular velocity profile for model m0.6 and
Fig. \ref{st0.6} displays the radius versus time profile for selected mass
zones.  
The negative gradient in rotational velocity is greatest at the
boundary between the original homologous core and the accreting matter
around 2 -- 3 $\times 10^{6}$ cm.  The strong negative gradient in angular
velocity drives the magnetorotational instability (MRI) dynamo, and this
region is expected to be associated with a strong magnetic field
\citep{aki03}.


\begin{figure}[htp]
\centering
\includegraphics[totalheight=0.4\textheight]{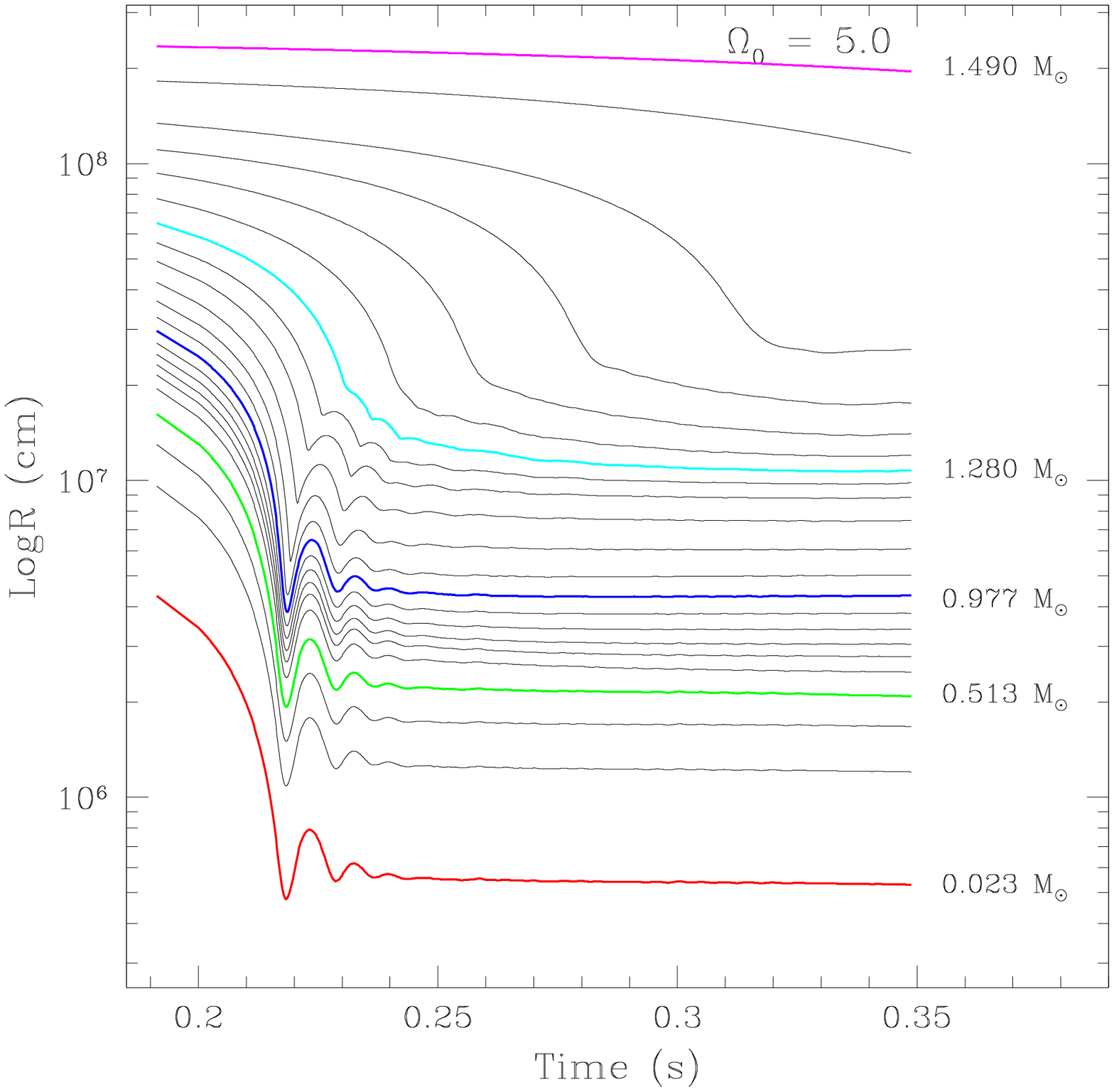}
\figcaption[stream5.0.eps]{The radius versus time of selected
mass zones for model m5.0.
  \label{st5.0}}
\end{figure}

Models m5.0, m6.0, and m7.0 bounce at sub--nuclear densities due to the
centrifugal potential.  Centrifugal force not only results in sub--nuclear
density bounce, but also delays the time to achieve bounce.  Model m8.0
did not bounce by the end of our computation.  Model m5.0, which is the
first model in the series we computed that bounces at sub--nuclear
density, bounced with a central density of $1.0 \times 10^{14}$ g
cm$^{-3}$.  As seen in Fig. \ref{st5.0}, the structure is significantly
less compact at bounce than models with lower initial rotation, and it
oscillates a few times after bounce.  This oscillation is undoubtedly an
artifact of our spherical calculation, but in these calculations it is
affected by the fact that the models that bounce at sub--nuclear density
have $\gamma \sim 4/3$ at the time of bounce up to a radius of $\sim$
10$^{7}$ cm. The condition of essentially neutral dynamical stability
makes them susceptible to radial perturbations with little energy penalty.
The structure of these rotationally--supported models contracts after
bounce while accreting mass from the outer layers.

\begin{figure}[htp]
\centering
\includegraphics[totalheight=0.4\textheight]{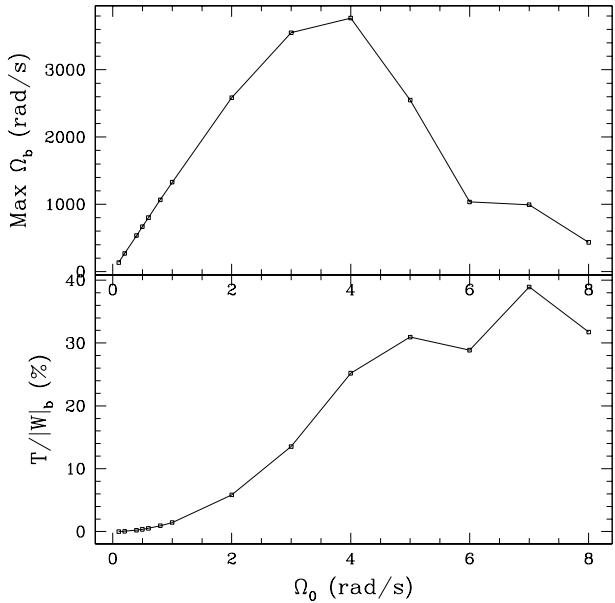}
\figcaption[omeome.eps]{Maximum rotational velocity and $T/|W|$ 
at bounce as a function of initial angular velocity.  \label{ome}}
\end{figure}

The maximum value of angular velocity occurs at the boundary of the
homologous core at bounce for models that bounce above nuclear density
(see Fig. \ref{ome0.6}); the locus of maximum angular velocity later
shifts to regions within the homologous core as the core relaxes after
bounce.  For sub--nuclear bounce models, the large scale angular velocity
gradient is negative throughout the structure, so the innermost region of
the core has the maximum value of angular velocity.  For modest initial
rotation, $\Omega_{0} < 3$ rad s$^{-1}$, 
the rotational velocity at bounce
increases with $\Omega_{0}$, as expected from \S 2.  More rapid rotation
gives a more significant centrifugal potential after bounce.
The peak in angular velocity is delayed in time for higher initial
rotation and the post--bounce angular velocity increases with time for the
most rapidly rotating models.  For initial rotation large enough that the
core bounces at sub--nuclear density due to the centrifugal effect, the
rotational velocity at bounce declines with increasing $\Omega_{0}$
because the core is not as compact at bounce.  In our calculations, the
peak of the rotational velocity at bounce corresponds to model m4.0, as
shown in Fig. \ref{ome}.

As the core experiences the collapse, the ratio $T/|W|$ increases (see
Fig. \ref{ome}) along with the increase in rotational velocity because
\begin{equation}
T \sim \frac{L^2}{MR^2},\ |W| \sim \frac{GM^2}{R},
\end{equation}
so that 
\begin{equation}
\label{twr}
\frac{T}{|W|} \sim \left( \frac{L^2}{GM^3} \right) \frac{1}{R}.
\end{equation}
The angular momentum is nearly conserved and no mass loss is expected up
to the core bounce, so that $L$ and $M$ are constants in our calculations
and eq. (\ref{twr}) is a function only of $R$.  Meanwhile, for a given
radius, the ratio $T/|W|$ is larger for bigger $L$.  While the rotational
velocity at bounce has a peak with increasing $\Omega_{0}$, $T/|W|$
continues to increase according to eq. \ref{twr} (Fig. \ref{ome}).  Among
our models that bounce at nuclear density, those with $\Omega_{0} = 3.0$
and $4.0$ s$^{-1}$ may be unstable to secular and dynamical instability,
respectively, since the $T/|W|$ values are close to $14\%$ and $27\%$,
respectively.  These models are expected to depart from spherical and even
axial symmetry, and our one--dimensional simulation is not adequate to
follow the dynamics after bounce quantitatively. The more rapidly rotating
models also have comparable or higher values of $T/|W|$ and hence should 
not be taken literally.

While the specifics of the non--monotonic behavior of the post--collapse
angular velocity shown in Fig. \ref{ome} undoubtedly depend on the
restrictions of our ``shellular" model, we suggest that the general
behavior will occur even in multidimensional models. We note that a peak
in angular velocity as a function of initial angular momentum is observed
in the two dimensional calculations of \citet{ott04} in their Fig. 19.  
The final angular velocity increases from their model with $\beta_{i} =
0.10\%$ to that with $\beta_{i} = 0.20\%$ ($\beta$ is their parameter for
the ratio of rotational to gravitational energy). For their model with
initial differential rotation, the final angular velocity begins to
decline with increasing initial angular momentum for their model with
$\beta_{i} = 0.50\%$.  Similarly, their model with initial solid body
rotation exhibits a peak in final rotational velocity for models with
$\beta_{i} = 0.50\% - 1.00\%$.  Our model with maximum rotation at bounce
(m4.0) corresponds to $T/|W|_{i} = 1.37\%$.  Direct comparison of our
models with those of \citet{ott04} is difficult since the simulations of
\citet{ott04} are two dimensional and do not include neutrino transport,
but their models clearly show a non--monotonic behavior of final angular
velocity as a function of initial angular momentum.  In our calculations,
models that bounce at sub--nuclear densities have larger $T/|W|$ than more
slowly rotating models that bounce at nuclear density, while the models in
\citet{ott04} show a peak in $T/|W|$ (their $\beta$) for the first
sub--nuclear bounce model at the time of bounce after which $T/|W|$
decreases with increasing initial rotation (see their Fig. 7). It seems
unlikely that the difference is our inclusion of neutrino transport since
the low densities associated with sub--nuclear bounce will mute the
neutrino production. Rather, we suspect the difference is that the
two--dimensional models allow some of the gravitational potential to be
converted into circulatory motion in addition to rotational motion whereas
our models force rotation on shells and hence a large contribution to $T$.  
We also neglect dissipation between shells in these trial calculations, an
effect that will also tend to maximize $T$.

Despite the limitations on our models with strong centrifugal force and
high $T/|W|$, the general trend seen in our calculations can be argued by
simple physics.  Starting from small initial angular velocity, the larger
the initial angular velocity, the larger the centrifugal force and angular
velocity at bounce.  When the centrifugal force is strong enough to halt
the collapse before the core achieves nuclear density, the angular
velocity is smaller than it would be otherwise because the core is not as
compact.  As the initial rotational velocity increases, the centrifugal
force is stronger, and core bounce occurs at lower density and later time.
This implies that there should be a peak at which the rotational velocity
is rapid, but the centrifugal force is not strong enough to yield
sub--nuclear bounce. Such a peak is observed in 
Fig. \ref{ome}, and happens near $\Omega_{0} = 4.0$ rad s$^{-1}$ in our
calculations.

\section{Summary and Conclusions}

There are several possible outcomes from core collapse that pertain to the
nature of the explosions and the nature of the compact remnants left
behind.  Core collapse can result in a successful or a failed supernova
explosion, and in the case of a successful explosion there are several
types of supernova (Type II, Type Ib, Type Ic, etc.), and some supernovae
are associated with gamma--ray bursts.  The central compact remnants can
be neutron stars or black holes (formed directly or by fall back) with a
range of rotation rates and magnetic field strengths.  Differences in
outcome with respect to compact remnants are usually attributed to the
mass of the progenitor core and of the envelope or to a combination of the
initial mass and the metallicity (through mass loss rates) because both
are key ingredients for massive star evolution, which determines the final
structure of the star.  In general the iron core mass shows an overall
trend to increase with increasing initial stellar mass, although there is
small scale non--monotonic variation of the relation between the iron core
and initial stellar mass \citep[see their Fig. 4]{bar90, tim96}.

\citet{heg03} presented a schematic diagram of the fate of compact objects
and of supernova types in terms of initial stellar mass and metallicity.  
For low metallicity stars, the compact objects are, from small to large
initial stellar mass, neutron star ($<$ 25 M$_{\odot}$) -- black hole by
fallback ($<$ 40 M$_{\odot}$) -- direct black hole ($<$ 140 M$_{\odot}$)
-- no remnant due to pair--instability supernova ($<$ 260 M$_{\odot}$) --
direct black hole ($>$ 260 M$_{\odot}$).  This scheme is modified for
stars with high mass loss rates, i.e. higher metallicity and high mass
($>$ 25 M$_{\odot}$), because high mass loss creates smaller helium cores
compared with stars of the same mass but with lower metallicity.  For
higher metallicity stars, the fate of the collapse becomes a function of
both the initial stellar mass and the metallicity.  For example, a star
with a mass of 60 M$_{\odot}$ is argued by \citet{heg03} to form a black
hole directly if the metallicity is low, but a black hole by fallback for
intermediate metallicity, and a neutron star for metallicity even larger
than the solar value.  For stars in the lower mass range (10 $<$
M$_{\odot}$ $<$ 25), mass loss is not severe, the fate of compact objects
does not depend on metallicity, and the outcome is predicted to always be
a neutron star.  Therefore, in terms of initial stellar mass and
metallicity, our model of a 15M$_{\odot}$ star should produce a neutron
star regardless of its metallicity.

There is a reasonable case to be made that rotation is involved with the
explosion mechanism due to circumstantial evidence from pre--collapse
(stellar evolution) and post--explosion (pulsar rotation) phases.  The
axisymmetric morphology of many supernova nebulae such as the rings of
SN~1987A is also considered to be associated with a rotational axis
\citep{wan02}.  Although hydrodynamical instabilities ($\ell$ = 1 mode)
alone can generate an axisymmetric asphericity \citep{fog02, blo03}, the
existence of rotation may inhibit the growth or otherwise modify the
nature of the instability \citep{blo03}.  Models without rotation cannot
explain pulsar activity, so core collapse models with some amount of
angular momentum are clearly required.  Despite the indication that
rotation accompanies core collapse, it is difficult to quantify the amount
of angular momentum in the pre--collapse iron core and in the newly born
proto neutron star (PNS) since current observations in electromagnetic
waves are not capable of revealing that quantity (except perhaps
indirectly by polarization).  Perhaps detection of gravitational waves
from core collapse supernovae will bring progress on this issue \citep[and
references therein]{ott04, mul04}.

We argue that rotation should be a third parameter along with the initial
stellar mass and metallicity that rules the nature of the explosion and
the fate of the compact objects. Rotation is an important ingredient in
stellar evolution where it can affect the mass loss rates, the nuclear
evolution through meridional mixing, and the collapse dynamics \citep[][and
references therein]{heg03,hir04}.  In our parametric study of rotating
core collapse models, we have shown that the resulting rotational velocity
of the PNS is very likely to be a non--monotonic function of initial
rotational velocity (Fig. \ref{ome}).  The peak of the maximum rotational
velocity naturally happens at the boundary separating conditions for which
the collapsed core bounces at over nuclear density from conditions
resulting in bounce at sub--nuclear density.  The model at the peak
post--bounce angular velocity possesses a large angular momentum with a
compact high--density core.  This non--monotonic behavior could have
consequences for the explosion and for the fate of the resulting compact
object.

We extend this reasoning to argue that the role of magnetic fields will
also depend non--monotonically on the initial iron core rotation.
\citet{aki03} demonstrated that the core collapse environment is prone to
the MRI, which amplifies seed magnetic fields to a strength of $\sim$
10$^{15}$ G.  \citet{kot04} and \citet{yam04} performed two--dimensional
MHD core collapse simulations, and their initial fields (10$^{9 - 12}$ G)
are amplified to $\sim 10^{13 - 16}$ G due to linear wrapping of the
magnetic fields from the attendant differential rotation.  Magnetic fields
of this strength can affect the explosion via viscous heating
\citep{tho05} and/or by dynamical MHD jets \citep{kot04,yam04}.  The
effects on the equation of state are estimated to be negligible nearby the
PNS where the density is high \citep[H. Duan 2004, private
communication;][]{aki04}, but if a highly magnetic bubble is convected to
a low density region, there could be important effects. There could also
be an effect on the neutrino cross section \citep{lai01a,bha02,dua04}.

If jets are necessary for an explosion to occur, then the possibility
exists that cores with low initial rotation fail to produce an explosion
and hence yield black holes, either directly or by fall--back if the
explosion is weak.  Peak post--bounce angular velocity could then
represent the region of maximum shear, maximum MRI dynamo, maximum affect
on neutrino transport and hence maximum explosion strength, guaranteeing a
left--over neutron star. Even higher initial rotation with less final
rotation and shear could yield weaker magnetic fields and other effects
and could yield black hole formation once again. Some fraction of the
latter could then yield gamma--ray bursts for more massive progenitor
stars.  The exact effects of this non--monotonic behavior could depend on
the mass and metallicity of the progenitor.  In this hypothetical scheme
it is not clear where magnetars arise.  We note that magnetars are
characterized by their large dipole fields that are some, possibly
complex, remnant of the post-bounce toroidal fields expected to be
generated by the MRI and other related effects.  We return to the origin
of magnetars below.


Although dynamical effects of magnetic fields are not included in this
study, corresponding magnetic field strengths were calculated for each
rotational model from m0.1 to m8.0.  We have adopted the fiducial
saturation model of \citet{aki03} using their eq. (8),
\begin{equation}
B_{sat}^{2} = 4 \pi \rho r^{2} \Omega^{2},
\end{equation}
with an exponential growth timescale of
\begin{equation}
\tau = 4 \pi \left| \frac{N^{2}}{2 \Omega} + \frac{d\Omega}{d\ln{r}} 
\right|^{-1}.
\end{equation}
This is the timescale of the maximum growing mode of the linear MRI for the 
case with entropy gradients \citep{aki03}.

In these models, the initially small seed fields are amplified near the
core bounce due to the strong shear generated by the rapidly rotating
compact PNS core and accreting material.  We note that the linear
stability analysis is local, however, and the MRI has not been established
in a dynamical background environment where the radial inflow velocity is
faster than the rotational velocity during core collapse.  Therefore the
evolution of the field strength before the core bounce may not reflect the
true effects of the MRI.  Once the core has halted collapse and accreting
matter inside the shock has a slow radial infall velocity, the MRI linear
stability analysis is valid \citep[see their Fig. 15]{aki03}; this
pertains after core bounce when the structure is in a quasi--steady state.
The field in a given mass zone can evolve somewhat after bounce even in
these dissipationless calculations as a given zone expands or contracts,
conserving flux.

\begin{figure}[htp]
\centering
\includegraphics[totalheight=0.4\textheight]{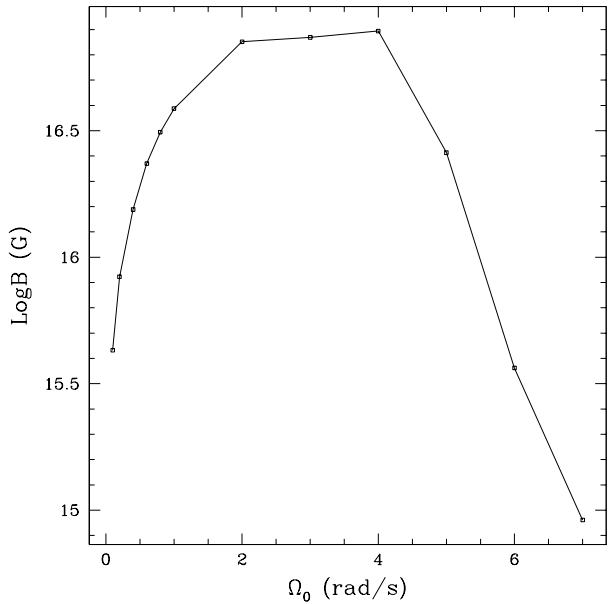}
\figcaption[maxblast.eps]{Magnetic field strengths for mass shells at
the boundary of the PNS core at $\sim$ 30 ms after bounce of the 
respective runs
as a function of initial angular velocity.  The
non--monotonicity in rotational velocity is reflected in the magnetic
field strength as well.  
\label{mag2}}
\end{figure}

The non--monotonic behavior of rotational velocity is also reflected in
the magnetic field strength.  The peak of the rotational velocity was
found to occur for model m4.0, which also corresponds to the highest
estimated value of the magnetic field at bounce.
The final saturation field strength grows with $\Omega_0$ up to model m3.0
and then decreases for the models with more rapid rotation that bounce at
sub--nuclear densities (Fig. \ref{mag2}).  For models m0.1 through m7.0,
the final saturation fields range between $10^{15} - 10^{17}$ G,
corresponding to a range of a factor of $10^4$ in quantities like the
magnetic pressure and hoop stress that depend on the square of the field
strength.  Even the slow rotating models reach magnetic field strengths of
order $10^{15}$ G in tens of milliseconds. The peak field strengths, $\sim
10^{17}$ G, are nearly large enough to be of direct dynamical importance.

Our simulations only ran at most to $\sim$ 500 msec after the beginning of
collapse. This is comparable to the expected explosion timescale, but we
note that there is a critical phase yet to come, the de--leptonization
phase, lasting perhaps 10 s, during which the rotating PNS will contract
and spin up substantially as it becomes a full-fledged neutron star.  
This phase could be a phase of renewed MHD vigor that we have not
explored. If collapse to a black hole does not occur during this phase,
then there is a renewed chance of generating even stronger magnetic fields
by means of the MRI and hence of driving MHD jets or other processes that
can affect the outcome of the supernova. Note that this possibility of
renewed MHD activity occurs as the potential for neutrinos to affect the
explosion diminishes.

If the supernova is successful, the de--leptonization phase will be
followed by a phase of neutron star spindown by magnetic dipole radiation
and gravitational wave radiation \citep{ost69} for the case of typical
neutron stars, and by magnetic braking for the case of magnetars
\citep{tho04}.  These are important epochs for the birth of neutron stars,
in which the transition is made from the fury of core collapse to the
neutron stars we observe today after the supernova has cleared.  The
critical nature of this transition phase motivates investigations of the
origin and evolution of magnetic fields and rotation \citep{tho93,tho01}
and their effects on neutrino--driven winds \citep{tho04}, on neutron star
recoil mechanisms \citep{lai01}, on nucleosynthesis yields, and more.

An important issue is how to make the connection between the rapidly
rotating models we describe here and the ``normal" pulsars that have
modest rotations and Crab-like dipole fields. While some pulsars may be
born rotating as fast as a few ms, others may be born rotating more
slowly. Examples are the 65 ms pulsar in G11.2-0.3, the remnant of SN 386
(Torii et al. 1997), and the 66 ms pulsar in 3C 58, the remnant of SN 1181
for which Murray et al.  (2002) estimate an initial spin of 60 ms. The
crucial issue is that the ``initial spin" for a pulsar astronomer is the
``final spin" for a supernova dynamicist, the spin after the supernova has
cleared away. In the jet-driven hypothesis, the rotation of the new-born
neutron star represents the fly--wheel that provides the rotational energy
to drive the (perhaps magnetically--catalyzed) jets. In this picture, the
supernova kinetic energy results in significant part by spinning down the
neutron star. Pulsars with relatively long ``initial" rotation periods
might then result from neutron stars that had done an especially efficient
job of dumping their rotational energy into the expansion energy of the
supernova ejecta.

One of the processes that could enhance the depletion of the angular
momentum of the neutron star is the formation of a bar with the attendant
generation of dynamic, MHD, and gravitational waves.  When differential
rotation is strong, the value of $T/|W|$ for onset of non--axisymmetric
instabilities is substantially decreased \citep{shi02,shi03,sai03,wat05}.  
\citet{ott05} showed that spiral and bar--like modes can develop in a
new--born PNS at $T/|W| \sim$ 0.08.  If the threshold for instability is
low, this mechanism may apply to a wide range of rotating core--collapse
models. For the current calculations (c.f. Fig. 5), all the models with
initial angular velocity in excess of about 3 s$^{-1}$, that is, all the
most rapidly rotating PNSs could be subject to these instabilities.
We note that the timescale for MRI--generated fields to saturate, $\sim$
10 ms, may be somewhat shorter than the time for the bar to fully develop,
$\sim$ 50 - 100 ms \citep{ott05}, so that the production of magnetic
fields and the growth of the bar would be substantially separate phases in
the evolution of a rapidly--rotating PNS. If a bar instability causes a
loss of angular momentum, then PNSs may spin down to just the angular
momentum at which they achieve stability. There may thus be a tendency for
pulsars to be born with just that angular momentum that represents the
critical value of $T/|W|$ for developing a bar mode. In practice, this
limit may be a function of PNS conditions, the de--leptonization phase
must also be considered, and other physical processes can contribute to
loss of angular momentum.  A great deal of work must be done to get to the
point where we can confirm or deny these various possibilities, but the
goal of connecting the rotational and magnetic condition of new-born
neutron stars to the observed properties of pulsars is of paramount
importance.

Another critical issue in the attempt to understand the rotational and
magnetic state of new-born neutron stars is the origin and evolution of
magnetars.  The field strengths we derive for these models are reminiscent
of, and even exceed, the fields associated with magnetars.  On the other
hand, it is very important to bear in mind that the fields generated by
the MRI will be strongly toroidal and will be maximal near the boundary of
the PNS. The fields associated with magnetars are the poloidal, dipole,
components of the neutron star thousands of years later.  Also, since the
angular velocity gradient is positive deep within the PNS at the time of
bounce, the resulting toroidal magnetic field profile is not anchored deep
inside the PNS. This means the large fields that the MRI can develop in
tens of milliseconds may be especially easy to shed as the surrounding
material dissipates in the supernovae explosion. We point out that the
magnetar dipole field may not be generated by the MRI, but generated
within the PNS core by an efficient $\alpha - \Omega$ dynamo as suggested
by \citet{tho93} after spin down of the PNS due to de--leptonization.  
The condition for an efficient $\alpha - \Omega$ dynamo requires rapid
rotation so that the Rossby number, the ratio of the rotation time to the
convective time, is small, nearly unity \citep{dun92,tho93}.  Because of
the non--monotonic dependence of the final rotational velocity on initial
iron core rotation, magnetar progenitors may not correspond to the most
rapidly rotating iron cores, but to iron cores with some intermediate
initial rotational velocity.  In addition, the condition to drive an
$\alpha - \Omega$ dynamo may not arise during the explosion of the
supernova, but during or after the contraction of the PNS due to
de--leptonization. At core bounce, model m4.0 has a rotational period of
$\sim 1.7$ msec at the boundary of the PNS core.  If we consider spin up
of the core by contraction at a later time, this model may be able to
provide the rapid rotation needed for an efficient $\alpha - \Omega$
dynamo, hence giving rise to magnetar magnetism.  The condition to produce
magnetar dipole fields, like the condition to maximize the rotation and
the field strength in a way to maximally promote a supernova explosion,
suggests that making an MHD jet-induced (or jet-aided) supernova and
producing a magnetar may require conditions near the peak in the
non-monotonic, post-collapse, rotation rates that we emphasize here.  The
fact that the most energetic supernova explosions may require maximal
effect of the MRI, and that magnetars may require maximal effect of the
$\alpha - \Omega$ dynamo means that the conditions for these sets of
outcomes may not be identical. Leaving behind a magnetar requires a
successful supernova explosion, but the opposite need not be true.  Thus,
the channel for the production of magnetars must be a subset of the
conditions that give successful supernovae.  Future study may reveal which
range of initial iron core rotation rates yields successful supernovae,
and which conditions within that range yield magnetars.

\acknowledgments
We thank I. Lichtenstadt for his generosity in supplying the code, and P.  
H\"{o}flich and E. L. Robinson for helpful discussion.  We also thank the
anonymous refree for useful comments.  This work was supported in part by
NASA Grant NAG59302 and NSF Grant AST-0098644.



\end{document}